\documentclass[floats,aps,prb,twocolumn,longbibliography,superscriptaddress]{revtex4-2}

\usepackage{url}
\usepackage{bm}
\usepackage{graphicx}
\usepackage{amsmath}
\usepackage{amstext}
\usepackage{amssymb}
\usepackage{amsfonts}
\usepackage{amsbsy}
\usepackage{verbatim}
\usepackage{xcolor}
\usepackage[colorlinks=true, urlcolor=blue, linkcolor=blue, citecolor=blue, pdftex]{hyperref}
\usepackage{multirow}
\usepackage{floatrow}
\usepackage{float}
\usepackage{gensymb}
\usepackage{textcomp}
\usepackage{enumitem}
\usepackage[version=4]{mhchem}
\usepackage{afterpage}
\usepackage{pbox}
\usepackage{makecell}
\usepackage{array,booktabs}
\usepackage{dsfont}
\usepackage{siunitx}
\usepackage[utf8]{inputenc}
\usepackage{braket}
\usepackage[normalem]{ulem}
\usepackage[version=4]{mhchem}

\begin{document}

\title{Evidence of rotational and tilting disorder of ReO$_6$ octahedra in single crystals of a 5\textit{d}$^1$ double perovskite Ba$_2$CaReO$_6$}

\author{Alasdair Nicholls}
\affiliation{Laboratory for Quantum Magnetism, Institute of Physics, \'Ecole Polytechnique F\'ed\'erale de Lausanne, CH-1015 Lausanne, Switzerland}

\author{Jian-Rui Soh}
\affiliation{Laboratory for Quantum Magnetism, Institute of Physics, \'Ecole Polytechnique F\'ed\'erale de Lausanne, CH-1015 Lausanne, Switzerland}
\affiliation{Quantum Innovation Centre (Q.InC), Agency for Science Technology and Research (A*STAR), 2 Fusionopolis Way, Singapore 138634}

\author{Yikai Yang}
\affiliation{Department of Engineering Science, University of Oxford, Parks Road, Oxford OX1 3PJ, United Kingdom}

\author{Alessandro Bombardi}
\affiliation{Diamond Light Source, STFC Rutherford Appleton Laboratory, Harwell Science and Innovation Campus, Didcot, Oxfordshire, OX11 0QX, UK\looseness=-1}

\author{Daigorou Hirai}
\affiliation{Department of Applied Physics, Nagoya University, Nagoya 464-8603, Japan}

\author{Henrik M. R{\o}nnow}
\affiliation{Laboratory for Quantum Magnetism, Institute of Physics, \'Ecole Polytechnique F\'ed\'erale de Lausanne, CH-1015 Lausanne, Switzerland}

\author{Ivica {\v Z}ivkovi{\' c}}
\email{ivica.zivkovic@epfl.ch}
\affiliation{Laboratory for Quantum Magnetism, Institute of Physics, \'Ecole Polytechnique F\'ed\'erale de Lausanne, CH-1015 Lausanne, Switzerland}

\date{\today}

\begin{abstract}
We present results of an experimental study on single crystals of a 5\textit{d}$^1$ double perovskite \ce{Ba2CaReO6}. Magnetization measurements reveal a weak splitting between zero-field-cooled and field-cooled protocols below 12\,K. At magnetic fields above 1\,T the splitting is absent and the magnetic susceptibility is featureless. A detailed specific heat study in a wide temperature range and comprising different heat pulses did not reveal any indication of a thermodynamic phase transition. At low temperatures we do observe specific heat deviating from a phonon background, leading to a total electronic entropy release of $\sim Rln2$. Resonant and non-resonant x-ray diffraction of characteristic Bragg peaks indicates a significant presence of disorder, potentially related to random tilts and rotations of rigid \ce{ReO6} octahedra.
\end{abstract}

\maketitle

\section{Introduction}

In recent years, correlated electron systems in the presence of a strong spin-orbit coupling (SOC) have attracted considerable attention~\cite{WitczakKrempa2014}, particularly with new quantum phases such as topologically-protected Weyl semi-metals~\cite{Liu2019a} and bond-anisotropic Kitaev spin-liquids~\cite{Takagi2019}. Among them, higher-rank multipole order is particularly intriguing, since it has often been termed as a `hidden order' due to the lack of direct detection of the order parameter, despite the evidence of symmetry-breaking phase transitions from thermodynamic measurements. This was particularly emphasized in \textit{f}-electron systems, like \ce{URu2Si2}~\cite{Kasuya1997}, \ce{UPd3}~\cite{McMorrow2001}  and \ce{NpO2}~\cite{Tokunaga2006}. Recent progress with 5\textit{d}-based double perovskites offers a novel venue to explore the properties of phases where interactions other than dipole-dipole dictate the ordering tendencies. 

The general formula for double perovskites $A_2$$BB'$O$_6$ comprises of two metal sites, which in the rock-salt configuration~\cite{Anderson1993} forms a regular, three-dimensional alternating pattern of \textit{B} and \textit{B'} ions. The multipolar order has been investigated in systems where \textit{A} and \textit{B} are non-magnetic cations while \textit{B'} is a heavy transition metal ion (Ta, W, Re, Os) with a \textit{d}$^1$ or \textit{d}$^2$ electron configuration~\cite{Chen2010,Chen2011}. Currently, strong experimental evidence indicates two phase transitions in the \textit{d}$^1$ systems \ce{Ba2MgReO6} and \ce{Ba2NaOsO6}, a charge antiferro-quadrupole order followed by magnetic dipole order with canted antiferromagnetic configuration at lower temperatures~\cite{Erickson2007,Lu2017,Hirai2019,Soh2024}. Those two compounds are available as high-quality single crystals, with other members of the \textit{d}$^1$ family, which have only been prepared as powders so far, following a similar trend~\cite{Marjerrison2016,Ishikawa2021,Barbosa2024}. A notable exception is \ce{Ba2LiOsO6} which exhibits only a single transition into an antiferromagnetic state~\cite{Barbosa2024}.

Recently, we managed to synthesize single crystals of \ce{Ba2CaReO6}, a compound which differs from \ce{Ba2MgReO6} only in the size of the \textit{B} cation ($r_\mathrm{Mg^{2+}} = 72 \text{\,\AA}$, $r_\mathrm{Ca^{2+}} = 100 \text{\,\AA}$)~\cite{Shannon1976}. A previous report using powder samples of \ce{Ba2CaReO6}~\cite{Ishikawa2021} indicates two phase transitions, but unlike \ce{Ba2MgReO6}, the magnetic dipoles order antiferromagnetically, with a field-induced crossover to a canted state above $~\sim 40$\,T. Here we show that the single crystal specimens differ substantially from their powder counterparts, lacking any evidence of thermodynamic transitions. This is potentially caused by the underlying structural disorder related to the rotation and tilting of \ce{ReO6} octahedra.

\section{Experimental details}

Single crystals were grown using the flux method. A mixture of \ce{BaCl2} and \ce{CaCl2} in a 2:1 molar ratio was used as a flux. Starting materials BaO, CaO, and \ce{ReO3} were mixed in a 2:1:1 molar ratio, combined with the flux in a 1:1 mass ratio, and sealed in a welded Pt tube with an inner diameter of 5\,mm and a length of 20\,mm, containing a total of 0.2\,g of starting materials and flux. The tube was heated to 1500$^\circ$C at a rate of 200$^\circ$C/h, then slowly cooled to 1050$^\circ$C over 20 hours, followed by furnace cooling. Afterward, the sample was extracted and thoroughly washed with distilled water to remove residual flux. The total mass of the obtained crystals was approximately 10\,mg. Crystals with a maximum diameter of 600\,$\mu$m were obtained, with an average grain size of about 200\,$\mu$m. Magnetization was measured in a commercial magnetometer using an extraction method and a Superconducting Quantum Interference Device (SQUID) with a magnetic field up to 7\,T (MPMS3, Quantum Design). Specific heat was measured in a commercial system using a relaxation method with magnetic fields up to 9\,T (PPMS, Quantum Design). Temperature-dependent resonant and non-resonant x-ray diffraction was performed at the I-16 beamline at Diamond Light Source, Didcot, UK, using a closed-cycle cryostat. Pyrolytic graphite (008) was used as an analyzer.

\section{Experimental Results}

\subsection{Magnetization}

\begin{figure}
\centering
\includegraphics[width=0.9\columnwidth]{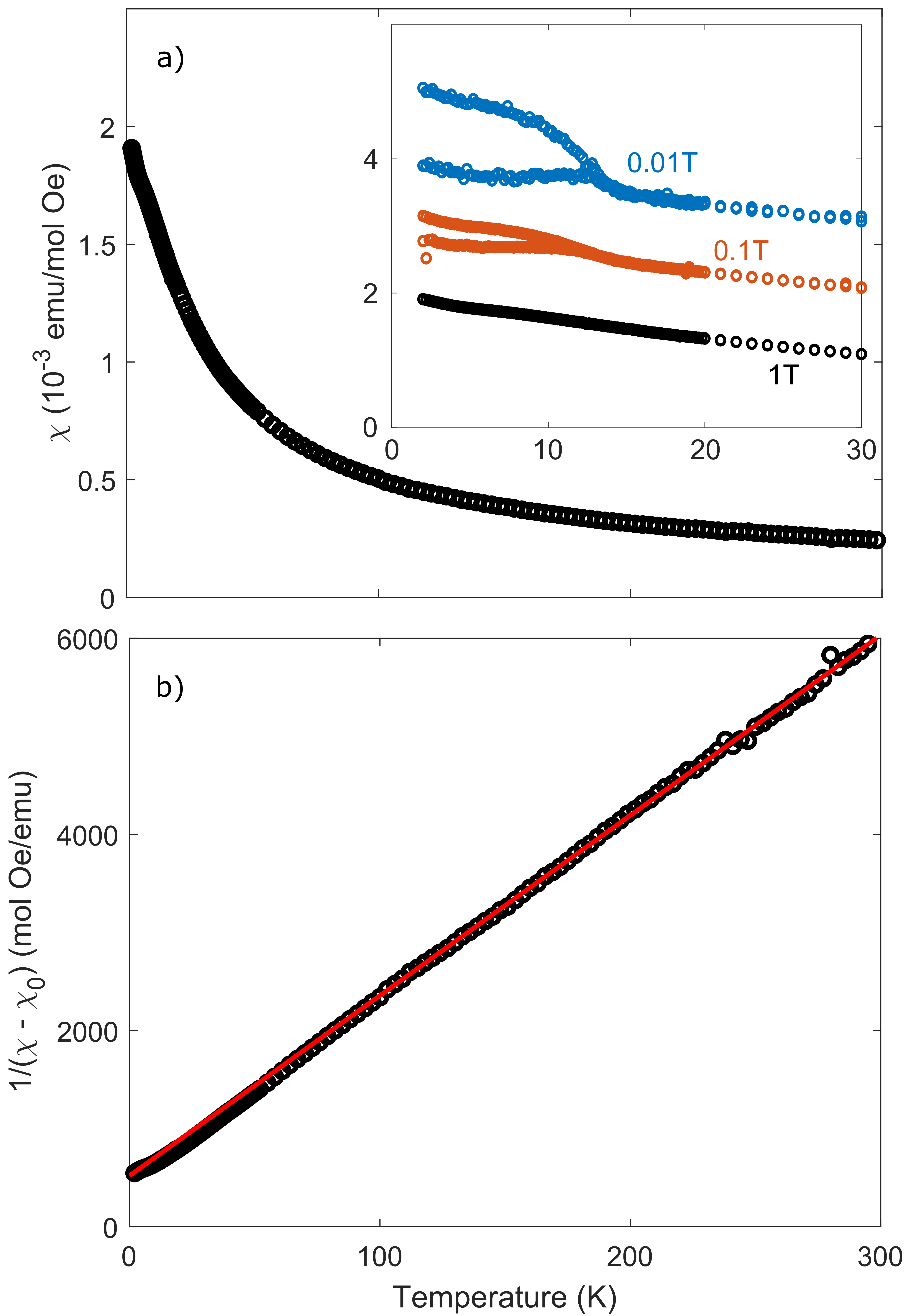}
\caption{(a) Susceptibility $\chi(T) = M/H$ of \ce{Ba2CaReO6} single crystals along [111] direction at $\mu_0H = 1$\,T. The inset shows the evolution of ZFC and FC protocols for different values of magnetic field. 0.01\,T and 0.1\,T measurements are shifted for clarity. (b) Temperature dependence of the inverse susceptibility $1/\chi$, alongside the Curie-Weiss behavior represented by the red straight line.}
\label{fig:magnetization}
\end{figure}

The magnetic susceptibility $\chi_\mathrm{d.c.} = M/H$ of \ce{Ba2CaReO6} single-crystals, measured at $B = 1$\,T and presented in Figure~\ref{fig:magnetization}a, monotonically grows as temperature decreases, without any features typically associated with long-range magnetic order. However, at smaller values of magnetic field, an evident splitting of zero-field cooled (ZFC) and field-cooled (FC) protocols is observed below 12\,K The splitting diminishes for larger magnetic field values (see inset of Figure~\ref{fig:magnetization}a). Above the splitting the susceptibility remains field-independent.

The inverse susceptibility, measured at $B = 1$\,T, is plotted in Figure~\ref{fig:magnetization}b, together with a result of a fit to the Curie-Weiss behavior $\chi = C/(T- \theta) + \chi_0$. In a wide temperature range (100\,K - 300\,K) we observe a good agreement, with $\theta = -28$\,K indicating a predominant antiferromagnetic coupling. The extracted Curie constant $C = 0.054$\, emu\,K/mol corresponds to an effective moment of $\mu_\mathrm{eff} \approx \sqrt{8*C} = 0.66$\,$\mu_\mathrm{B}$. Similar values have also been found previously in powder sample of \ce{Ba2CaReO6}~\cite{Ishikawa2021}, as well as in \ce{Ba2MgReO6}~\cite{Hirai2019}. We note that a small deviation from the Curie-Weiss behavior could be seen below 100\,K, similar to previously reported observations~\cite{Ishikawa2021}.





\begin{figure}
\centering
\includegraphics[width=0.8\columnwidth]{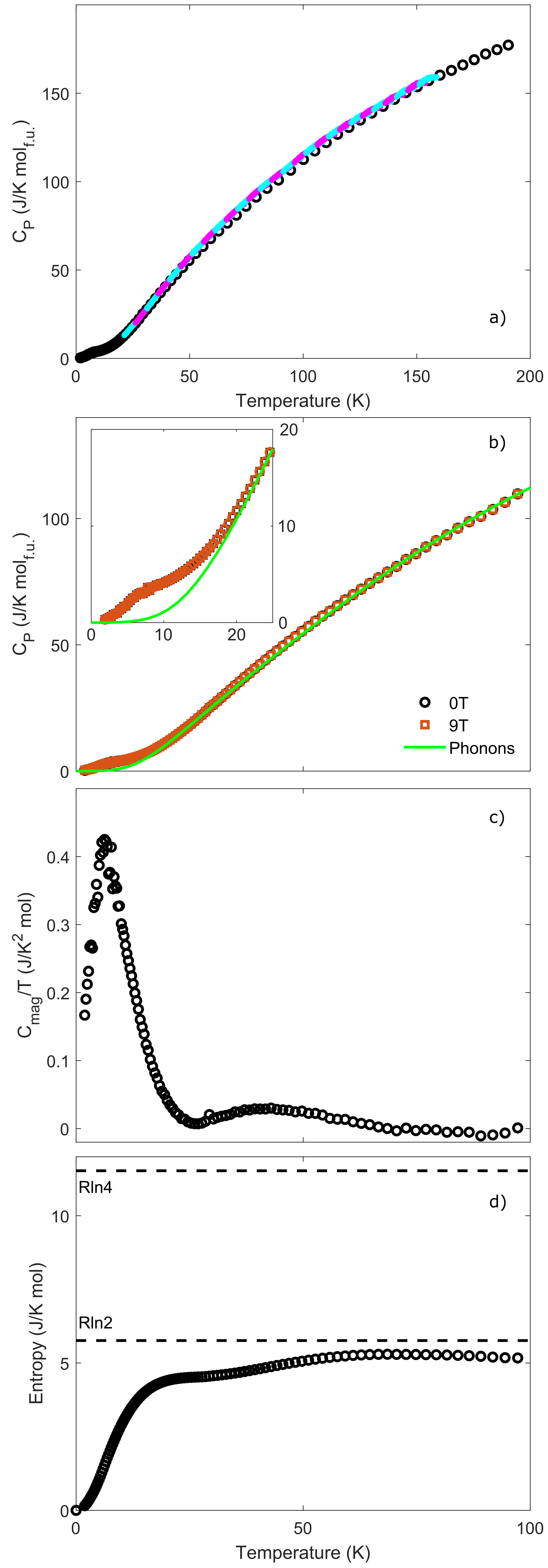}
\caption{(a) Specific heat of \ce{Ba2CaReO6} using short (black circles) and long heat pulses (colored dashed line). (b) Specific heat of \ce{Ba2CaReO6} single crystals in a magnetic field of 0\,T and 9\,T, alongside an estimated phonon background from \textit{ab-initio} calculations, which has been scaled along both \textit{T}- and \textit{$C_P$} axes to match the experimental data in the range 80 -- 100\,K~\cite{Zivkovic2024}. The inset zooms in at the low temperature range where a clear deviation from a purely phonon contribution is observed. (c) Specific heat divided by temperature after a subtraction of the phonon contribution. (d) Temperature dependence of entropy related to electronic degrees of freedom.}
\label{fig:SpecificHeat}
\end{figure}

\subsection{Specific heat}

Specific heat of \ce{Ba2CaReO6} has been measured in a wide temperature range, Figure~\ref{fig:SpecificHeat}a. At low temperature, below 20\,K, we see a deviation from a purely phononic contribution, here approximated with previously published DFT calculations for \ce{Ba2MgReO6}~\cite{Pasztorova2023,Zivkovic2024}. In the first approximation both Ca$^{2+}$ and Mg$^{2+}$ ions act as electron donors, without establishing covalent bonds with oxygen ions. The renormalization of specific heat is therefore proportional to a ratio of molar masses $\sqrt{M_\mathrm{\ce{Ba2CaReO6}}/M_\mathrm{\ce{Ba2MgReO6}}}$ but at the same time the effect of Ca and Mg mass difference is strongly diminished by much larger masses of barium and rhenium ions. Given that DFT calculations themselves have a small deviation from experimental results obtained for \ce{Ba2MgReO6}~\cite{Pasztorova2023}, we argue that the phonon background indicated in Figure~\ref{fig:SpecificHeat}b is a good starting point for the discussion of electronic specific heat of \ce{Ba2CaReO6}.

Figure~\ref{fig:SpecificHeat}c shows the result of subtracting the phonon contribution from the total measured specific heat. The deviation below 20\,K exhibits a peak around 6\,K in $C/T$, with an additional broad contribution centered around 40\,K and extending up to 70\,K. We note that the latter might arise as a consequence of an improper phonon subtraction. However, as can be seen in Figure~\ref{fig:SpecificHeat}d, the majority of entropy seems to be concentrated at temperatures below 20\,K. With both contributions taken into account, the total integrated entropy reaches just below $Rln 2$, the result recently found in \ce{Ba2MgReO6}~\cite{Zivkovic2024} to be a reflection of a dynamic Jahn-Teller effect.

\begin{figure}
\centering
\includegraphics[width=0.9\columnwidth]{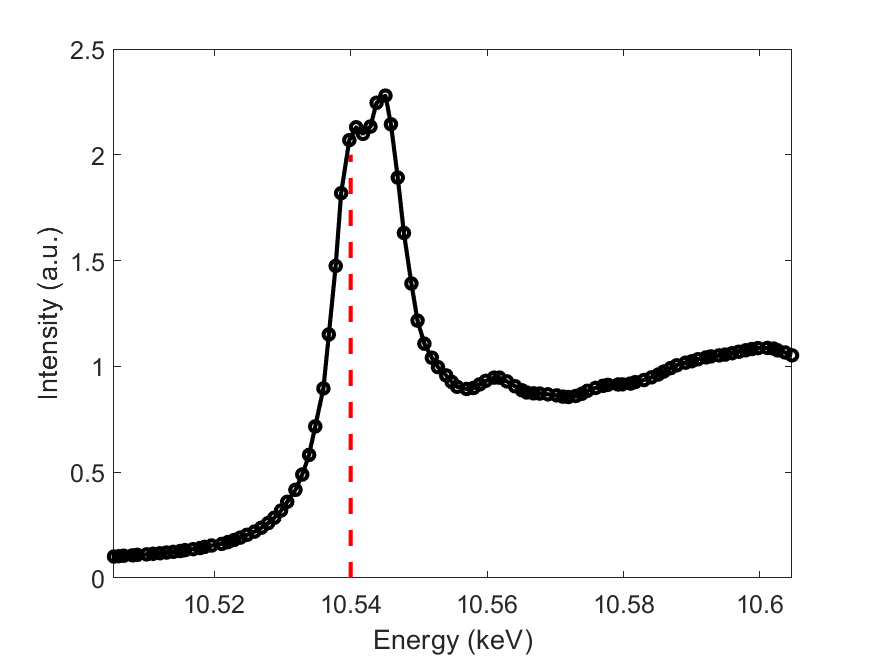}
\caption{Fluorescence of single-crystal \ce{Ba2CaReO6} sample around the Re \textit{L}$_3$ edge. The vertical dashed line indicates the resonant energy used in REXS experiments.}
\label{fig:Fluorescence}
\end{figure}

\begin{figure}
\centering
\includegraphics[width=0.9\columnwidth]{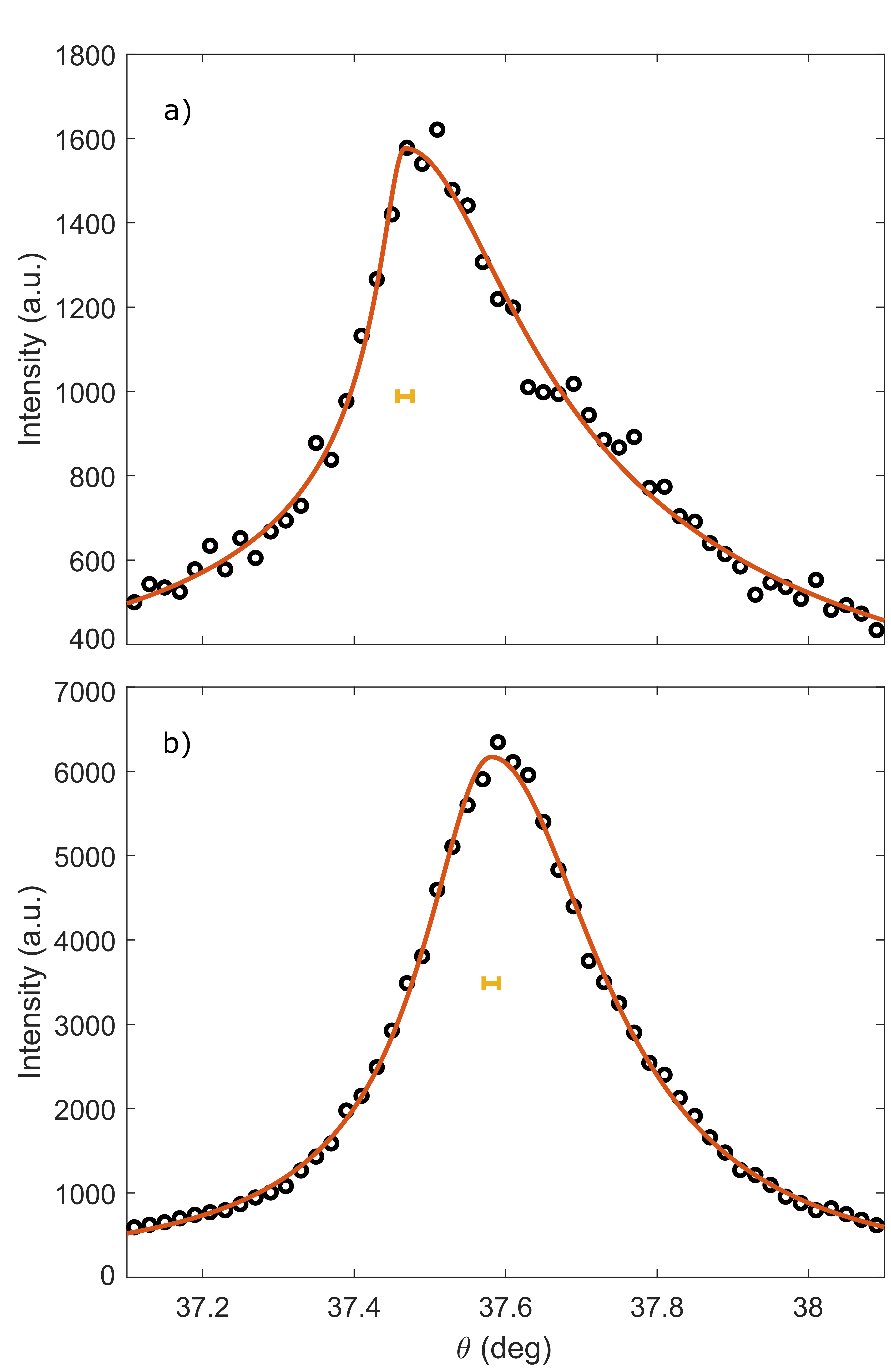}
\caption{A $\theta$-scan at the [0,10,0] Bragg peak at (a) 300\,K and (b) 10\,K in the $\pi \sigma'$ channel. The horizontal segment indicates the width of a typical structural peak at I-16. The solid lines represent the best fit using an asymmetric Pearson VII curve.}
\label{fig:zero10zero300K}
\end{figure}

The deviation from the Curie-Weiss behavior observed around 100\,K has been attributed previously to a tetragonal compression, accompanied with a splitting of a ground state $j_\mathrm{eff} = 3/2$ quartet into two doublets~\cite{Ishikawa2021}. Corresponding features have been observed in specific heat and thermal expansion measurements~\cite{Ishikawa2021}. In our case, specific heat results do not indicate any deviation from a smooth background that can be associated with the phonon contribution. To verify our results, alongside the typical data obtained in a step-wise manner using small (1-2\%) heating pulses, we also performed measurements using long heat pulses, amounting to 50\% or more of the base temperature, and overlapping with each other (the dashed colored line in Figure~\ref{fig:SpecificHeat}a). In this case specific heat is extracted from $dT/dt$, allowing to observe any weak features, including first-order phase transitions, that might have been missed or simply absorbed by the short heat pulse procedure. In a wide temperature range, extending well above temperatures previously reported ($\sim 130$\,K)~\cite{Ishikawa2021}, the two methods lead to the same result. We note that the `missing entropy' of an additional $Rln 2$ at higher temperatures ($T \geq 100$\,K) would have to come from a substantial contribution to $C_P$ due to the temperature denominator in the expression $S_\mathrm{mag} = \int{C_P/T dT}$, and therefore would be rather visible on top of a smooth, phonon-based background.





\subsection{X-ray diffraction}


In order to look for microscopic evidence of long-range order, we employed resonant elastic x-ray scattering (REXS) at the Re \textit{L}$_3$-edge. The energy dependence of the fluorescence signal from the single crystal of \ce{Ba2CaReO6} is presented in Figure~\ref{fig:Fluorescence}. It exhibits a double-peak structure, similar to what has been reported in \ce{Ba2MgReO6}~\cite{Soh2024,Frontini2024}. With the energy set to the first maximum of the fluorescence ($E = 10.540$\,keV), we searched for evidence of long-range magnetic order at low temperatures ($T = 7$\,K). At [5,5,0], a structurally forbidden peak which was observed in \ce{Ba2MgReO6}~\cite{Soh2024} as an indicator of antiferromagnetic order, we could not find any trace of scattering in the $\sigma \pi^\prime$ channel.

Alternatively, we looked at [0,10,0] in the $\pi\sigma^\prime$ channel, a structural Bragg peak on top of which any ferromagnetic component of long-range order could be traced in \ce{Ba2MgReO6}~\cite{Soh2024}. Here, we observed that already at room temperature the scattering signal acquires a width significantly larger than the experimental resolution typical for I-16, Figure~\ref{fig:zero10zero300K}. The peak was visibly asymmetric which we tried to phenomenologically describe using an asymmetric (split) Pearson VII curve. A Pearson VII curve is defined by: 

\begin{equation}
    P(x, A, x_0, m, \sigma) = \frac{A}{\left(1 + \left[\frac{2(x - x_0) (2^{\frac{1}{m}}-1)^{\frac{1}{2}})}{\sigma}\right]^2\right)^m}
\end{equation}

where $A$ describes the height of the peak, $x_0$ is the centre of the peak, $\sigma$ is the full width at half maximum and $m$ is the function shape parameter. The shape parameter $m$ can take any value in the range $[0, \infty]$ where $m = 1$ describes a Lorentzian, whilst $m\rightarrow\infty$ describes a Gaussian. We described the asymmetric peak shape with two Pearson VII functions, a `left' and `right' function based around $x_0$ to best capture the asymmetric nature of these peaks: 

\begin{equation}
\begin{split}
    P_L \equiv P(x, A, x_0, m_L, \sigma_L) \\
    P_R \equiv P(x, A, x_0, m_R, \sigma_R)
\end{split}    
\end{equation}

\begin{figure}
\centering
\includegraphics[width=0.9\columnwidth]{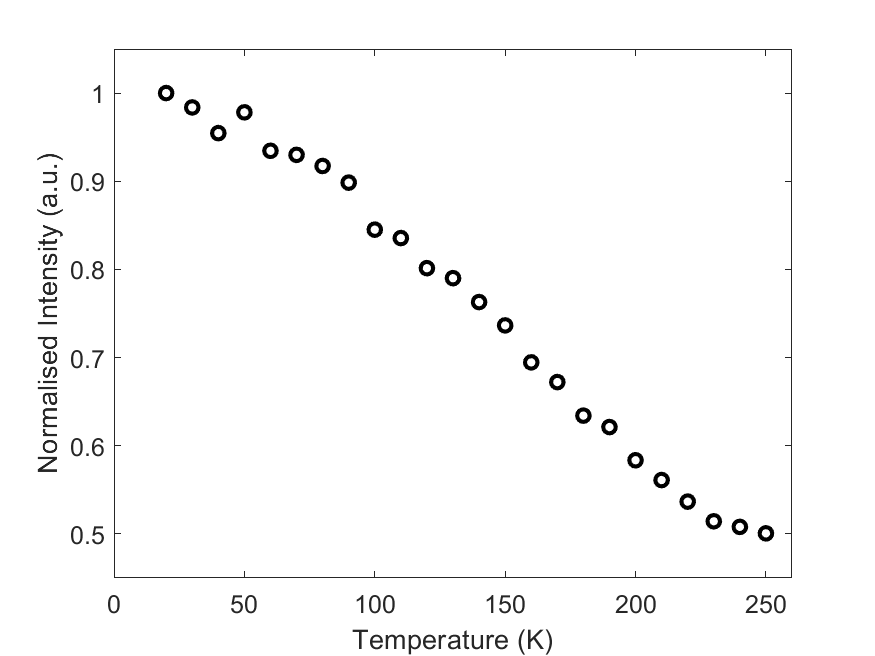}
\caption{Temperature dependence of the [0,10,0] Bragg peak intensity.}
\label{fig:Polarization}
\end{figure}

The intensity of the [0,10,0] peak gradually increases towards low temperatures, without any feature that would indicate a phase transition (Figure~\ref{fig:Polarization}).

\begin{figure}
\centering
\includegraphics[width=0.9\columnwidth]{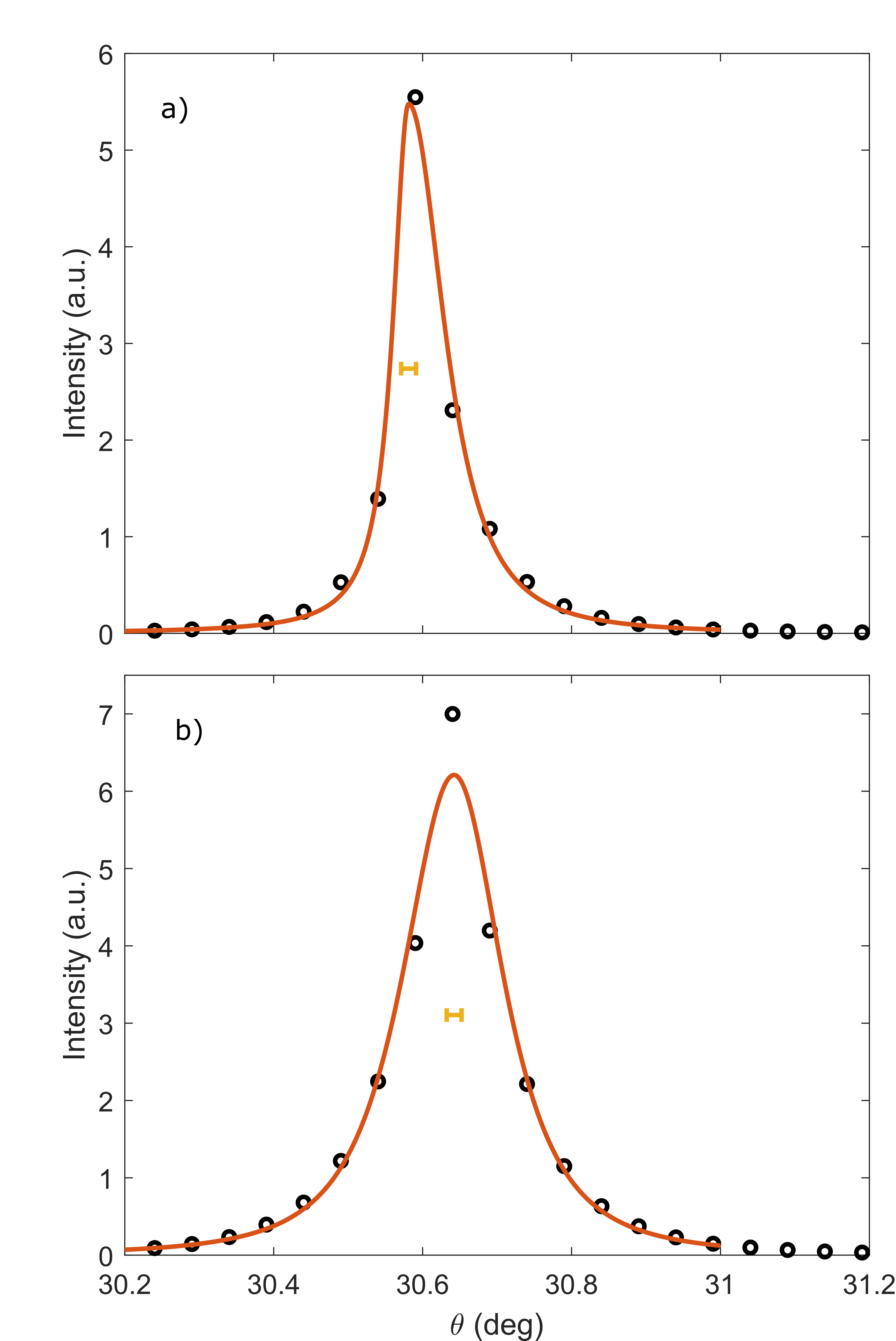}
\caption{A $\theta$-scan at the [4,4,0] Bragg peak at (a) 300\,K and (b) 10\,K in the $\sigma \sigma'$ channel. The horizontal segment indicates the width of a typical structural peak at I-16. The solid lines represent the best fit using an asymmetric Pearson VII curve.}
\label{fig:44zero280K}
\end{figure}

\begin{figure}
\centering
\includegraphics[width=0.9\columnwidth]{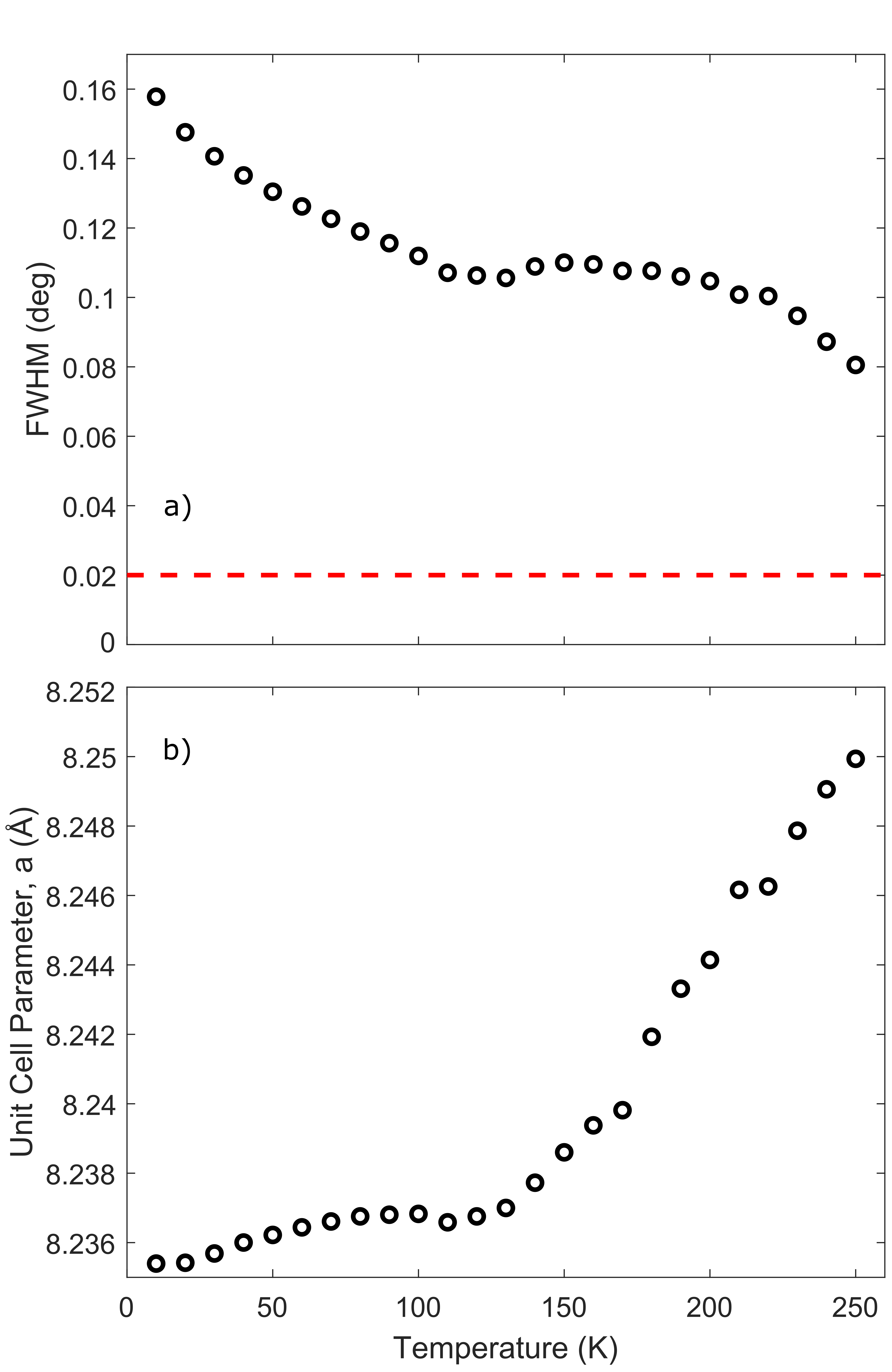}
\caption{Temperature dependence of (a) the extracted width of the [4,4,0] Bragg peak. The red dashed line represents a typical FWHM at I-16. (b) Temperature dependence of the unit cell parameter $a$ calculated from the position of the [4,4,0] Bragg peak.}
\label{fig:Broadening440}
\end{figure}

To verify the origin of the [0,10,0] peak broadening, we looked at the behavior of a purely structural peak in the off-resonance conditions ($E = 10.500$\,keV, $\sigma \sigma'$ channel). In Figure~\ref{fig:44zero280K} we plot a theta scan around the [4,4,0] peak and observe a similar broadening already around room temperature. The peak is further widened at low temperature, confirming the observation seen for [0,10,0]. The temperature dependence of the width of the [4,4,0] peak is non-monotonous, without an indication of a structural phase transition, Figure~\ref{fig:Broadening440}.

\begin{figure}
\centering
\includegraphics[width=0.9\columnwidth]{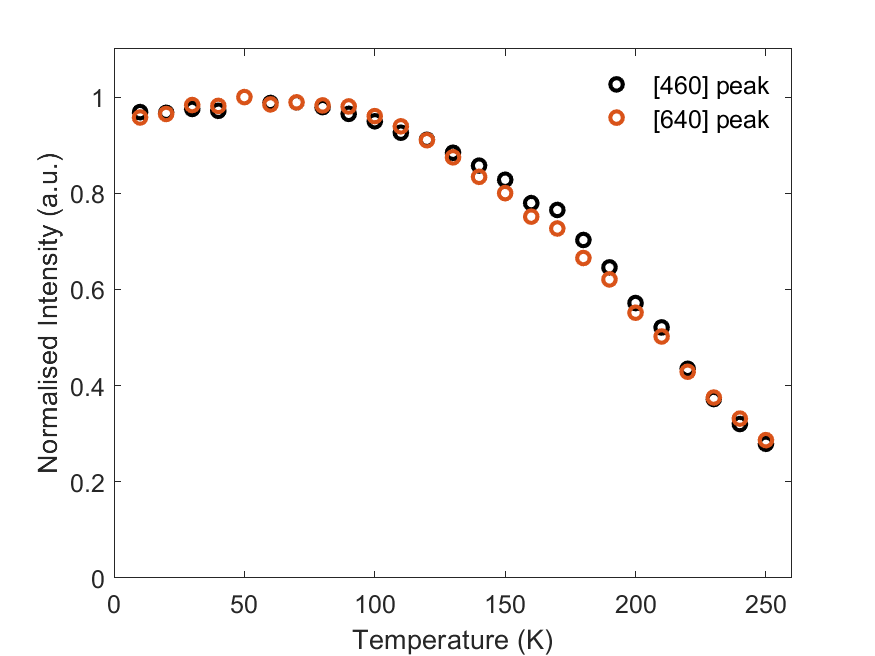}
\caption{Temperature dependence of the [4,6,0] and [6,4,0] Bragg peak intensity measured in the $\sigma \sigma'$ channel.}
\label{fig:Tilting}
\end{figure}

These results lead us to conclude that a certain amount of structural disorder is present in the material. A significant temperature dependence of the peak widths indicates that the disorder might not be arising from fixed types of defects, like vacancies and inclusions. Rather, it is potentially linked to structural deformations of unit cells through rotations and/or tilts. As can be seen from Figure~\ref{fig:Tilting}, the intensity of [6,4,0] and [4,6,0] structural peaks associated with these modifications is small, but non-zero, at room temperature and grows significantly towards low temperatures. Similarly, if we calculate the unit cell parameter within the cubic \textit{F}m$\overline{3}$m setting from the position of the [4,4,0] peak, we observe that a monotonic decrease persists down to 150\,K, below which a small upturn is observed and a gradual saturation occurs (Figure~\ref{fig:Broadening440}).

\section{Discussion}

Our set of data, comprising both thermodynamic and x-ray diffraction results, strongly indicate that single crystals of \ce{Ba2CaReO6} substantially differ from their powder counterparts~\cite{Ishikawa2021}. We see no evidence of a thermodynamic phase transition in specific heat, although a substantial entropy, related to degrees of freedom other than phonons, is released up to 80\,K, The weak ZFC/FC splitting seen in magnetization is hard to associate with long-range order, since no evidence of magnetic Bragg peaks appears at low temperatures and the splitting is completely washed out with magnetic field of 1\,T. We note that the magnetic susceptibility $\chi$ at low temperatures reaches values of $\sim 2 \cdot 10^{-3}$\,emu/mol Oe, a value characteristic of antiferromagnets and glassy-type systems. We suggest that the splitting comes from freezing of short-range correlations between magnetic moments, associated with the majority of electronic entropy being released below 20\,K.

The source of disorder is potentially linked to random tilts and rotations of individual \ce{ReO6} octahedra, as indicated from a substantial temperature dependence of [4,6,0] and [6,4,0] structural peaks. A recent resonant inelastic x-ray scattering on \ce{Ba2CaReO6} revealed~\cite{iwahara2025} the excitation spectra very similar to \ce{Ba2MgReO6}~\cite{Zivkovic2024}, indicating rigid \ce{ReO6} octahedra. It is worth noting that in \ce{Sr2MgReO6}, where a smaller Sr ion is replacing Ba, a coherent rotation of octahedra is observed, resulting in a tetragonal unit cell~\cite{Gao2020}. In \ce{Ba2CaReO6} a larger Ca ion is replacing Mg but at a different lattice site than in the case of \ce{Sr2MgReO6}. In the double perovskite structure the ligands are not shared between metal centers, allowing for each \ce{ReO6} octahedron to be independent of its neighbors. It is possible that the size of Ca ions is pushing the structure towards a critical value of the Goldschmidt tolerance factor~\cite{Goldschmidt1926} which coupled with a presence of usual random structural defects leads towards the observed behaviour. Nevertheless, further studies are needed to establish the exact microscopic origin of different behavior in powder and single crystal specimen of \ce{Ba2CaReO6}.

\section{Acknowledgements}

J.-R. S. acknowledges support from the Singapore National Science Scholarship, Agency for Science Technology and Research. I. {\v Z}., J.-R. S. and H.M.R. acknowledge support from the ERC Synergy grant HERO (Grant ID: 810451). H. M. R. acknowledges the support from SNSF Projects No. 200020-188648 and 206021-189644. The support from Diamond Light Source through the beamtime proposal number MM34820-1 at I-16 is acknowledged.

\bibliography{referencesBMRO.bib}

\end{document}